\documentclass[12pt,epsf]{article}

\usepackage{graphicx}

\newcommand{\be}{\begin{equation}}

\newcommand{\ee}{\end{equation}}

\newcommand{\ba}{\begin{array}}

\newcommand{\ea}{\end{array}}

\begin{document}
\begin{titlepage}
\vspace{.5in}
\begin{flushright}
CQUeST-2006-0026
\end{flushright}
\vspace{0.5cm}

\begin{center}
{\Large\bf The false vacuum bubble nucleation due to a nonminimally coupled scalar field}\\
\vspace{.4in}

  {$\rm{Wonwoo \,\, Lee}^{\S}$}\footnote{\it email:warrior@sogang.ac.kr}\,\,
  {$\rm{Bum-Hoon \,\, Lee}^{\dag\S}$}\footnote{\it email:bhl@sogang.ac.kr}\,\,
  {$\rm{Chul \,\, H. \, Lee}^{\ddag}$}\footnote{\it email:chulhoon@hanyang.ac.kr}\,\,
  {$\rm{Chanyong \,\, Park}^{\S}$}\footnote{\it email:cyong21@sogang.ac.kr}\\

  {\small \dag \it Department of Physics, Sogang University, 121-742, Seoul,
  Korea}\\
  {\small \S \it Center for Quantum Spacetime, Sogang University, 121-742, Seoul,
  Korea}\\
  {\small \ddag \it  Department of Physics, Hanyang University, 133-791,
  Seoul, Korea}\\

\vspace{.5in}
\end{center}
\begin{center}
{\large\bf Abstract}
\end{center}
\begin{center}
\begin{minipage}{4.75in}

{\small \,\,\,\, We study the possibility of forming the false
vacuum bubble nucleated within the true vacuum background via the
true-to-false vacuum phase transition in curved spacetime. We
consider a semiclassical Euclidean bubble in the Einstein theory
of gravity with a nonminimally coupled scalar field. In this paper
we present the numerical computations as well as the approximate
analytical computations. We mention the evolution of the false
vacuum bubble after nucleation.}

\end{minipage}
\end{center}
\end{titlepage}

\newpage
\section{ Introduction \label{sec1}}
What did the spacetime look like in the very early universe?
Probably there was a dynamical spacetime foam structure which was
introduced by John A. Wheeler, indicating that quantum
fluctuations come into play at the Planck scale, changing topology
and metric \cite{wheeler}. Also the cosmological constant, as a
dynamical variable rather than a universal constant \cite{brown},
may have played an important role as an ingredient which caused a
dynamical spacetime foam structure in the very early universe. But
such a phenomenon is difficult to describe in the theory of
gravity. On the other hand, there were investigations on the
mechanism of creating the inflationary universe in the laboratory
\cite{farhi}. However there is no regular method for the
nucleation of small regions of false vacuum. In this paper we show
that such complicated vacuum structure (spacetime structure) or
false vacuum bubble can occur by the vacuum-to-vacuum phase
transition in the semiclassical approximation.

It has been shown that the first-order vacuum phase transitions
occur via the nucleation of true vacuum bubble at zero temperature
both in the absence of gravity \cite{col1} and in the presence of
gravity \cite{col2}. This result was extended by Parke
\cite{parke} to the case of arbitrary vacuum energy densities. An
extension of this theory to the case of non-zero temperatures has
been found by Linde \cite{linde1} in the absence of gravity, where
one should look for the $O(3)$-symmetric solution due to
periodicity in the time direction $\beta$ with period $T^{-1}$
unlike the $O(4)$-symmetric solution in the zero temperature.
These processes as cosmological applications of false vacuum decay
have been applied to various inflationary universe scenarios by
many authors \cite{guth}.

As for the false vacuum bubble formation, Lee and Weinberg
\cite{lee1} have shown that if the vacuum energies are greater
than zero, gravitational effects make it possible for bubbles of a
higher-energy false vacuum to nucleate and expand within the true
vacuum bubble in the de Sitter space which has a topology of
4-sphere. The false vacuum bubble nucleation is described as the
inverse process of the true vacuum bubble nucleation. However,
their solution is larger than the true vacuum horizon \cite{basu}.
The oscillating bounce solutions, another type of Euclidean
solutions, have been studied in detail by Hackworth and Weinberg
\cite{hackworth}. On the other hand Kim et al.\ \cite{kim1} have
shown that false vacuum region may nucleate within the true vacuum
bubble as global monopole bubble in the high temperature limit.

In this paper we present that the false vacuum bubble can be
nucleated within the true vacuum background in the Einstein theory
of gravity with a nonminimally coupled scalar field. The
nonminimal coupling between a scalar field and gravity has been
discussed in various cosmological scenarios such as inflation
\cite{inf1} and quintessence \cite{quin1}. The nonminimally
coupled scalar field was introduced by Chernikov and Tagirov in
the context of radiation problems \cite{cher}. Other works with a
nonminimal coupling term are discussed in Ref.\ \cite{callan},
\cite{faraoni} and references therein.

In the semiclassical approximation, the vacuum-to-vacuum phase
transition rate per unit time per unit volume is given by
\begin{equation}
\Gamma/V = A e^{-B/\hbar}[1+ {\cal O}(\hbar)],
\end{equation}
where the pre-exponential factor $A$ is discussed in Refs.\
\cite{cal1} and the exponent $B$ is the Euclidean action. The
standard approach to the calculation of bubble nucleation rates
during the first order phase transition is based on the work of
Langer in statistical physics \cite{langer}, and a theory of the
decay of the false vacuum in spontaneously broken theories at zero
temperature has been first suggested by Voloshin, Kobzarev, and
Okun \cite{voloshin}. The standard results obtained by Coleman-De
Luccia and Parke were extended to the case of the nonminimal
coupling term \cite{lee2}, where the influence on the true vacuum
bubble radius and the nucleation rates was evaluated.

How can the false vacuum bubble be nucleated within the true
vacuum background via the true-to-false vacuum phase transition in
the context of gravity theory? The mechanism is not known in
curved spacetime with arbitrary vacuum energy in the pure Einstein
theory of gravity. In this work, we study the possibility of the
false vacuum bubble nucleation due to the nonminimal coupling of
the scalar field $\Phi$ to the Ricci curvature using Coleman-De
Luccia's semiclassical instanton approximation. We present the
numerical results as well as the approximate analytical
computations.

The plan of this paper is as follows. In section 2 we present the
formalism for the false vacuum bubble nucleation within the true
vacuum background in the Einstein theory of gravity with a
nonminimally coupled scalar field and our main idea for this work.
In section 3 the numerical solutions are obtained by solving the
Euclidean equation of motion of a scalar field in our model. Our
solutions represent the nucleation of small regions of false
vacuum bubble in curved spacetime with arbitrary vacuum energies.
In section 4 the exponent $B$ and the radius of the false vacuum
bubble are obtained analytically by employing Coleman's thin-wall
approximation in cases both of the false vacuum bubble nucleation
within the true vacuum background and of the true vacuum bubble
nucleation within the false vacuum background. In addition, we
analyze the evolution of the bubble after its nucleation. The
results are discussed.

\section{The false vacuum bubble nucleation in the Einstein theory of gravity with a nonminimally coupled scalar field}

In this section, we summarize the basic mechanism following the
work presented in Ref.\ \cite{lee2} with the correction of an
error term there and present the formalism for the nucleation of
false vacuum bubble within the true vacuum background with a
nonminimally coupled scalar field in our model. For this theory,
the action is given by
\begin{equation}
S = \int \sqrt{g} \: d^4x  \left[ \frac{R}{2\kappa} -
\frac{1}{2}{\nabla^\alpha}\Phi {\nabla_\alpha}\Phi - \frac{1}{2}
\xi R \Phi^2 - U(\Phi) \right] +S_b , \label{action}
\end{equation}
where $\kappa \equiv 8\pi G$, $g\equiv -det g_{\mu\nu}$, $U(\Phi)$
is the scalar field potential, $R$ denotes the Ricci curvature of
spacetime, the term $-\xi R\Phi^2 /2$ describes the nonminimal
coupling of the field $\Phi$ to the Ricci curvature and $\xi$ is a
dimensionless coupling constant. Hereafter we will omit the
boundary term \cite{gibbons}, $S_b$, because it will be cancelled
in these processes.

\begin{figure}[t]
\begin{center}
\includegraphics[width=4.5in]{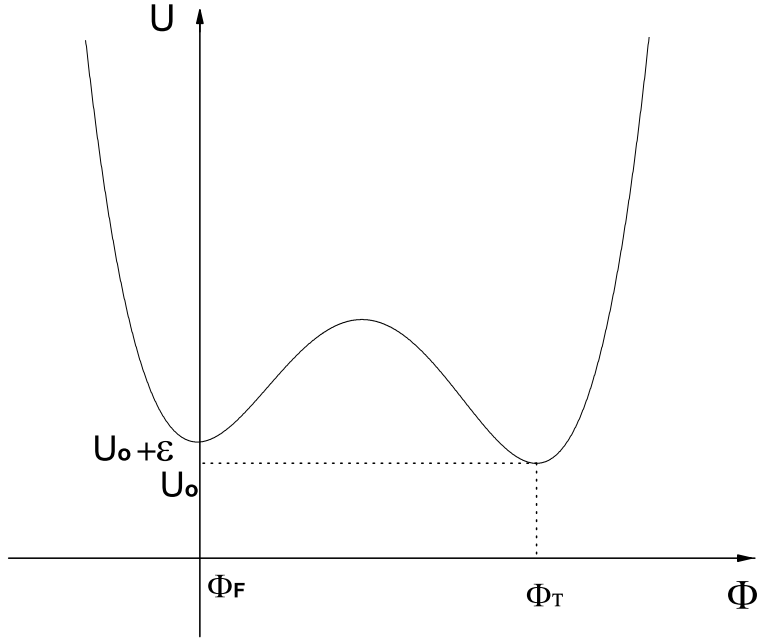}
\end{center}
\caption{\footnotesize{Potential with two minima,
$U(\phi_F)=U_o+\epsilon$ and $U(\phi_T)=U_o$.}} \label{fig:fig1}
\end{figure}

Let us consider the case where $U(\Phi)$ has the form
\begin{equation}
U(\Phi) = \frac{\lambda}{8} \Phi^2(\Phi-2b)^2 -
\frac{\epsilon}{2b}(\Phi-2b)+U_o ,
\end{equation}
where $\lambda$, $\epsilon$ and $b$ are positive parameters. The
minimum of the potential plays the role of the cosmological
constant term, and the potential $U(\Phi)$ has two nondegenerate
minima, one of which is lower than the other. $U(\Phi_T)$
corresponds to the true vacuum state and $U(\Phi_F)$ to the false
vacuum state, separated by a potential barrier (Fig.\
\ref{fig:fig1}). These vacuum states will be modified by the
nonminimally coupled scalar field, as is further discussed later.

The corresponding equation satisfied by the scalar field is
written by
\begin{equation}
\frac{1}{\sqrt{g}} \partial_{\mu} [\sqrt{g}
g^{\mu\nu}\partial_{\nu}\Phi] - \xi R\Phi - \frac{\partial
U}{\partial\Phi} =0.
\end{equation}

The Einstein equations are
\begin{equation}
R_{\mu\nu} - \frac{1}{2}g_{\mu\nu}R=\kappa T_{\mu\nu},
\end{equation}
where $R_{\mu\nu}$ is the Ricci tensor and $T_{\mu\nu}$ is the
matter energy momentum tensor,
\begin{eqnarray}
T_{\mu\nu} = \frac{1}{1-\xi\Phi^2 \kappa} \left[ \nabla_{\mu}\Phi
\nabla_{\nu}\Phi-g_{\mu\nu}\left(\frac{1}{2}\nabla^{\alpha}\Phi
\nabla_{\alpha}\Phi +U(\Phi)\right) \right. \nonumber \\
\left. + \xi(g_{\mu\nu} \nabla^{\alpha}
\nabla_{\alpha}\Phi^2-\nabla_{\mu} \nabla_{\nu}\Phi^2 )\right].
\end{eqnarray}
The curvature scalar is given by
\begin{equation}
R = \frac{\kappa[4U(\Phi)+ \nabla^{\mu}\Phi \nabla_{\mu}\Phi
-3\xi\nabla^{\mu}\nabla_{\mu}\Phi^2]}{1-\xi\Phi^2 \kappa}.
\end{equation}
Here we adopt the notations and sign conventions of Misner,
Thorne, and Wheeler \cite{misner}.

$O(4)$ symmetric bubbles have the minimum Euclidean action in the
absence of gravity \cite{col3} and seem to be a reasonable
assumption in the presence of gravity. In our work, we assume the
$O(4)$ symmetry for both $\Phi$ and the spacetime metric
$g_{\mu\nu}$ in a similar manner. The most general rotationally
invariant Euclidean metric is
\begin{equation}
ds^2 = d\eta^2 + \rho^2(\eta)[d\chi^2 +
\sin^2\chi(d\theta^2+\sin^2\theta d\phi^2)].  \label{gemetric}
\end{equation}
Then $\Phi$ is a function of $\eta$ only and one has $R_E
=-6(\rho'^2+\rho\rho''-1)/\rho^2$. The Euclidean action becomes
\begin{equation}
S_E = 2\pi^2 \int \rho^3(\eta) d\eta \left[ \frac{3}{\kappa}
\left\{ \left( \frac{\rho'}{\rho}\right)^2+\left(
\frac{\rho''}{\rho}\right)-\left(\frac{1}{\rho}\right)^2 \right
\}(1-\xi\Phi^2\kappa)+\frac{1}{2}\Phi'^2 + U(\Phi) \right],
\label{eaction}
\end{equation}
where the prime denotes the differentiation with respect to
$\eta$. The Euclidean field equations for $\Phi$ and $\rho$ turn
out to be
\begin{eqnarray}
 \Phi'' &+& \frac{3\rho'}{\rho}\Phi'-\xi R_E \Phi=\frac{dU}{d\Phi}, \label{ephi} \\
 \rho'^{2} &=&
 1+\frac{\kappa\rho^2}{3(1-\xi\Phi^2\kappa)}(\frac{1}{2}\Phi'^{2}-U),
 \label{erho}
\end{eqnarray}
respectively. The boundary conditions for the bounce are
\begin{equation}
\lim_{\eta \rightarrow \eta_{max}} \Phi(\eta) = \Phi_{T}, \,\,\,\,
 \frac{d\Phi}{d\eta}\Big|_{\eta=0}=0.
\end{equation}
where $\eta_{max}$ is a finite value in Euclidean de Sitter space
and $\eta_{max}=\infty$ in both Euclidean flat and anti-de Sitter
space.

\begin{figure}[t]
\begin{center}
\includegraphics[width=4.5in]{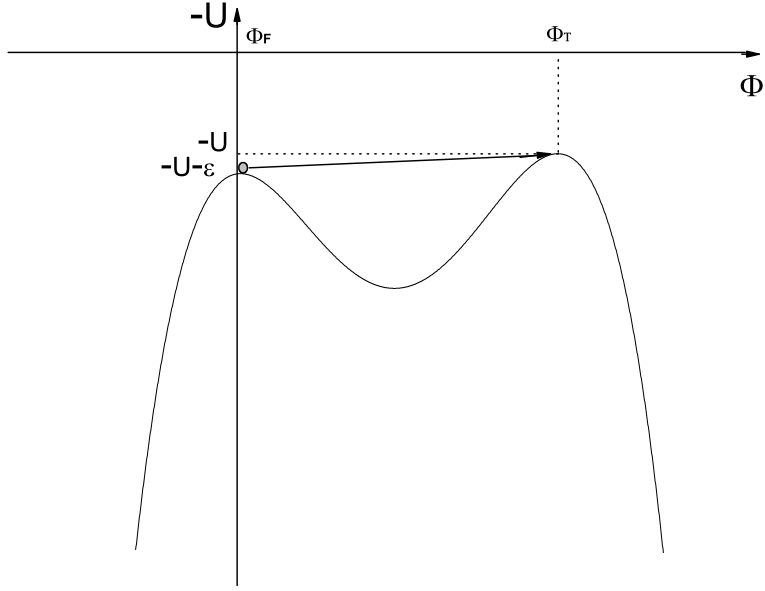}
\end{center}
\caption{\footnotesize{One-particle dynamics in the potential $-U$
.}} \label{fig:fig2}
\end{figure}

Eq.\ (\ref{ephi}) can be treated as a one-particle equation of
motion with $\eta$ playing the role of time in the corresponding
potential well, $-U(\Phi)$ (Fig.\ \ref{fig:fig2}). Multiplying
Eq.\ (\ref{ephi}) by $\frac{d\Phi}{d\eta}$ and rearranging the
terms, one obtains
\begin{equation}
 \frac{d}{d\eta}\left[\frac{1}{2}\Phi'^{2}-U \right]
 =-\frac{3\rho'}{\rho}\Phi'^{2} + \xi R_E \Phi\Phi'.
\end{equation}
The quantity in the square brackets here can be interpreted as the
total energy of the particle with the potential energy $-U$,
 the first term on the right hand side as the dissipation rate of the
total energy and the second term as the extra source of the power
which appears due to coupling to gravity. The second term, as the
extra power, plays an important role for the particle to reach at
$-U(\Phi_T)$. The true vacuum bubble nucleation within the false
vacuum background corresponds to the particle starting at some
point near $\Phi_T$ at $\eta=0$ and reaching $\Phi_F$ at
$\eta=\eta_{max}$. On the other hand, the false vacuum bubble
nucleation within the true vacuum background corresponds to the
particle starting at some point near $\Phi_F$ at $\eta=0$ and
reaching $\Phi_T$ at $\eta=\eta_{max}$. It will be possible that
the false vacuum bubbles nucleate within the true vacuum
background in the theory of gravity with such a term. The main
idea is simple. It will happen if and only if such a term could
overcome the second term on the left hand side of Eq.\
(\ref{ephi}) during the phase transition,
\begin{equation}
\xi R_E \Phi > \frac{3\rho'}{\rho}\Phi'.
\end{equation}
The second term on the left hand side of Eq.\ (\ref{ephi}) is
interpreted as a viscous damping term both in Euclidean flat and
anti-de Sitter space. In Euclidean de Sitter space, the term is
interpreted as a viscous damping term from $\rho=0$ to
$\rho=\rho_{max}$ and an accelerating term from $\rho=\rho_{max}$
to $\rho=0$ because the coefficient becomes negative. To
understand the role of the third term on the left hand side of
Eq.\ (\ref{ephi}), we can absorb $\Phi'$-independent term into the
effective potential. After the curvature scalar is substituted in
Eq.\ (\ref{ephi}) the field equation becomes
\begin{eqnarray}
{\Phi''} + \frac{3{\rho'}}{{\rho}}\Phi' -
\frac{\xi(1-6\xi){\kappa}{\Phi'}^2{\Phi}}{(1-\xi{\Phi}^2{\kappa}+6\xi^2{\Phi}^2{\kappa})}
&=& \frac{1-\xi{\Phi}^2{\kappa}}{(1-\xi{\Phi}^2{\kappa}+6\xi^2{\Phi}^2{\kappa})}\frac{d{U}}{d{\Phi}}+\frac{4\xi{\Phi}{\kappa} {U}}{(1-\xi{\Phi}^2{\kappa}+6\xi^2{\Phi}^2{\kappa})}, \nonumber \\
&=& \frac{d{U}_{eff}}{d{\Phi}}. \label{ueff}
\end{eqnarray}
The third term on the left hand side of Eq.\ (\ref{ueff}) is
nonlinear in $\Phi'$, and can be interpreted as an accelerating
term or viscous damping term depend on the signature.

The explicit form of the effective potential as well as the
position of the vacuum states of the effective potential is not
easy to write down in Eq.\ (\ref{ueff}). The position of the false
and true vacuum of $U(\Phi)$, $\Phi_F$ and $\Phi_T$, are not the
same as that of $U_{eff}(\Phi)$, $\Phi^{eff}_F$ and
$\Phi^{eff}_T$, because the position are dependent on
$\xi$-coupling. While the magnitude of the effective potential for
false and true vacuum, $U_{eff}(\Phi^{eff}_F)$ and
$U_{eff}(\Phi^{eff}_T)$, are not the same as that of $U(\Phi_F)$
and $U(\Phi_T)$ because the magnitude are also dependent on the
$\xi$-coupling. The nucleation of the false vacuum bubble for
$U(\Phi)$ may be understood as the "true vacuum" bubble nucleation
of the effective potential $U_{eff}(\Phi)$ via Coleman's mechanism
for bubble nucleation.

In this work we consider five particular cases; (Case 1) from de
Sitter space which has the true vacuum state of the lower positive
energy density to de Sitter space which has the false vacuum state
of the higher positive energy density, (Case 2) from flat space to
de Sitter space, (Case 3) from anti-de Sitter space to de Sitter
space, (Case 4) from anti-de Sitter space to flat space and (Case
5) from anti-de Sitter space which has the true vacuum state of
the higher negative energy density to anti-de Sitter space which
has the false vacuum state of the lower negative energy density.

In case 1, well inside the bubble where $\Phi$ remains constant at
$\Phi_F$ and $U=U_o+\epsilon$, the solution for $\rho$ is
$\rho=\Lambda\sin\frac{\eta}{\Lambda}$ and the metric is given by
\begin{equation}
ds^2 = d\eta^2 + \Lambda^2 \sin^2 \frac{\eta}{\Lambda} \{d\chi^2 +
\sin^2\chi (d\theta^2 +\sin^2\theta d\phi^2)\}, \label{inmetric-1}
\end{equation}
where $\Lambda= (\frac{3}{\kappa(U_o +\epsilon)})^{1/2}$. In the
region outside the bubble where $\Phi$ remains constant at
$\Phi_T$ and $U=U_o$, the solution for $\rho$ is $\rho=\Lambda_1
\sin \frac{\eta+\delta}{\Lambda_1}$ and the metric is given by
\begin{equation}
ds^2 = d\eta^2 + \Lambda_1^2 \sin^2 \frac{\eta+\delta}{\Lambda_1}
\{ d\chi^2 + \sin^2\chi (d\theta^2 +\sin^2\theta d\phi^2)\},
\label{outmetric-1}
\end{equation}
where $\Lambda_1= (\frac{3(1-4b^2\xi \kappa)}{\kappa U_o})^{1/2}$.
Notice that a constant $\delta$ is introduced so that $\rho$
inside can be continuously matched at the wall to $\rho$ outside.

In case 2, well inside the bubble, the metric is the same form as
Eq.\ (\ref{inmetric-1}). In the region outside, $\rho=\eta+\delta$
and the metric in this region is given by
\begin{equation}
ds^2 = d\eta^2 + (\eta+\delta)^2 \{ d\chi^2 + \sin^2\chi
(d\theta^2 +\sin^2\theta d\phi^2)\}. \label{outmetric-2}
\end{equation}

In case 3, well inside the bubble, the metric is also the same
form as Eq.\ (\ref{inmetric-1}) and outside the bubble, the
solution for $\rho$ is $\rho=\Lambda_2 \sinh
\frac{\eta+\delta}{\Lambda_2}$ and the metric is given by
\begin{equation}
ds^2 = d\eta^2 + \Lambda_2^2 \sinh^2 \frac{\eta+\delta}{\Lambda_2}
\{ d\chi^2 + \sin^2\chi (d\theta^2 +\sin^2\theta d\phi^2)\}.
\label{outmetric-3}
\end{equation}
where $\Lambda_2= (\frac{3(1-4b^2\xi \kappa)}{\kappa
|U_o|})^{1/2}$.

In case 4, the metric inside the bubble is given by
\begin{equation}
ds^2 = d\eta^2 + \eta^2 \{ d\chi^2 + \sin^2\chi (d\theta^2
+\sin^2\theta d\phi^2)\}, \label{inmetric-4}
\end{equation}
and outside the bubble, the metric is the same form as Eq.\
(\ref{outmetric-3}).

In case 5, the metric inside the wall is given by
\begin{equation}
ds^2 = d\eta^2 + \Lambda_3^2 \sinh^2 \frac{\eta}{\Lambda_3} \{
d\chi^2 + \sin^2\chi (d\theta^2 +\sin^2\theta d\phi^2)\},
\label{outmetric-4}
\end{equation}
where $\Lambda_3= (\frac{3}{\kappa (|U_o|-\epsilon)})^{1/2}$ and
the metric outside the bubble is the same form as Eq.\
(\ref{outmetric-3}).

\section{Numerical calculation}

The Euclidean field equations, (\ref{ephi}) and (\ref{erho}), can
be solved analytically. Hence in this section we solve the
equations numerically. We first rewrite the equations in terms of
dimensionless variables as follows
\begin{equation}
\frac{\lambda U(\Phi)}{\mu^4} = \tilde{U}(\tilde{\Phi}),\;\;
\frac{\lambda \Phi^2}{\mu^2}=\tilde{\Phi^2},\;\; \frac{\lambda
\epsilon}{\mu^4}=\tilde{\epsilon},\;\; \mu\eta=\tilde{\eta},\;\;
\mu\rho=\tilde{\rho}.
\end{equation}
These variables give
\begin{equation}
\tilde{U}(\tilde{\Phi})=\frac{1}{8}\tilde{\Phi}^2(\tilde{\Phi}-2)^2
- \frac{\tilde{\epsilon}}{2}(\tilde{\Phi}-2)+ \tilde{U_o},
\end{equation}
and the Euclidean field equations for $\Phi$ and $\rho$ become
\begin{eqnarray}
\tilde{\Phi''} &+&
\frac{3\tilde{\rho'}}{\tilde{\rho}}\tilde{\Phi'} -
\xi\tilde{R}_E\tilde{\Phi} = \frac{d\tilde{U}}{d\tilde{\Phi}}, \\
\tilde{\rho'}^{2} &=&
1+\frac{\tilde{\kappa}\tilde{\rho}^2}{3(1-\xi\tilde{\Phi}^2\tilde{\kappa})}(\frac{1}{2}\tilde{\Phi'}^{2}-\tilde{U}),
\end{eqnarray}
respectively, where $\tilde{R}_E=R_E/\mu^2$,
$\tilde{\rho}=\rho\mu$ and
$\tilde{\kappa}=\frac{\mu^2}{\lambda}\kappa$. The boundary
conditions also become
\begin{equation}
\lim_{\tilde{\eta} \rightarrow \tilde\eta_{max}}
\tilde{\Phi}(\tilde{\eta}) = \tilde{\Phi}_{T}, \,\,\,\,
 \frac{d\tilde{\Phi}}{d\tilde{\eta}}\Big|_{\tilde{\eta}=0}=0.
\end{equation}

\begin{figure}[t]
\begin{center}
\includegraphics[width=4.5in]{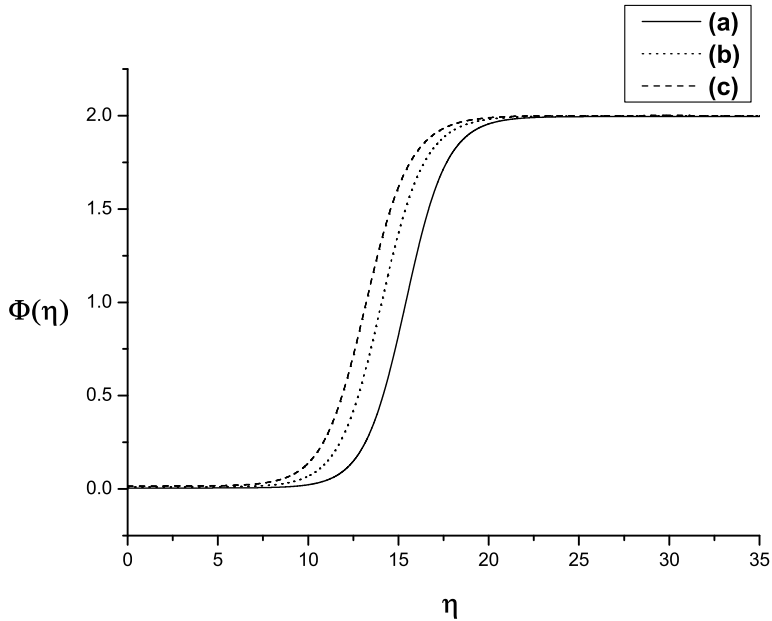}
\end{center}
\caption{\footnotesize{The false vacuum bubble profiles for
several values of $\tilde{\epsilon}$ and $\xi$ in case 1. Here
$\xi$ is taken to be positive value. The three curves are (a)
solid curve: $\tilde{\epsilon}=0.01$ and $\xi \simeq 0.271$, (b)
dotted curve: $\tilde{\epsilon}=0.02$ and $\xi \simeq 0.313$, (c)
dashed curve: $\tilde{\epsilon}=0.03$ and $\xi \simeq 0.346$.
Hereafter the axes denote the tilde attached variables.}}
\label{fig:fig3}
\end{figure}

In this work we consider the five particular cases of
true-to-false vacuum phase transitions; (Case 1) from de Sitter
space to de Sitter space, (Case 2) from flat space to de Sitter
space, (Case 3) from anti-de Sitter space to de Sitter space,
(Case 4) from anti-de Sitter space to flat space and (Case 5) from
anti-de Sitter space to anti-de Sitter space.

\begin{figure}[t]
\begin{center}
\includegraphics[width=4.5in]{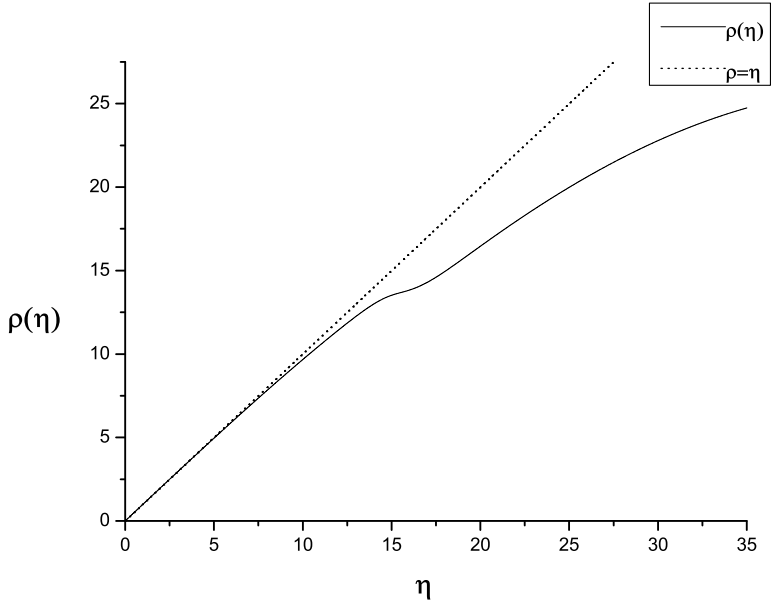}
\end{center}
\caption{\footnotesize{The evolution of
$\tilde{\rho}(\tilde{\eta})$ in case 1. The solid curve is the
solution of $\tilde{\rho}$ with $\tilde{\epsilon} =0.01$. In the
region inside the bubble, $\rho=\Lambda \sin
\frac{\eta}{\Lambda}$, and outside the bubble, $\rho=\Lambda_1
\sin\frac{\eta+\delta}{\Lambda_1}$.}} \label{fig:fig4}
\end{figure}

In Case 1, a scalar field originally in the true vacuum state of
the lower positive energy density decays into the false vacuum
state of the higher positive energy density. We have computed
three cases of bubble profiles corresponding to (a) solid curve:
$\tilde{\epsilon}=0.01$ and $\xi \simeq 0.271$, (b) dotted curve:
$\tilde{\epsilon}=0.02$ and $\xi \simeq 0.313$, (c) dashed curve:
$\tilde{\epsilon}=0.03$ and $\xi \simeq 0.346$ and we take
$\tilde{\kappa}=0.3, \tilde{U_T}=0.01$. We can see from the Fig.\
\ref{fig:fig3} that the radius of bubble becomes larger as
$\tilde{\epsilon}$ becomes smaller. But in order to climb the
hill, $U_{eff}(\Phi^{eff}_T)$, the value of $\xi$ becomes larger
as $\tilde{\epsilon}$ becomes larger. This can be understood
because $\xi$-term acts as an accelerating term which helps the
particle to climb the hill, up to $U_{eff}(\Phi^{eff}_T )$. The
evolution of $\tilde{\rho}(\tilde{\eta})$ is shown in Fig.\
\ref{fig:fig4}. The solid curve is the solution of $\tilde{\rho}$
with $\tilde{\epsilon} =0.01$. In the region inside the bubble,
$\rho=\Lambda \sin \frac{\eta}{\Lambda}$, and outside the bubble,
$\rho=\Lambda_1 \sin\frac{\eta+\delta}{\Lambda_1}$. Here we obtain
the small region of the false vacuum bubble within the true vacuum
background which is de Sitter space. The bending part of the solid
curve corresponds to the bubble wall. The thick bubble wall exists
when the difference in energy density between the false and true
vacuum is large.

\begin{figure}[t]
\begin{center}
\includegraphics[width=4.5in]{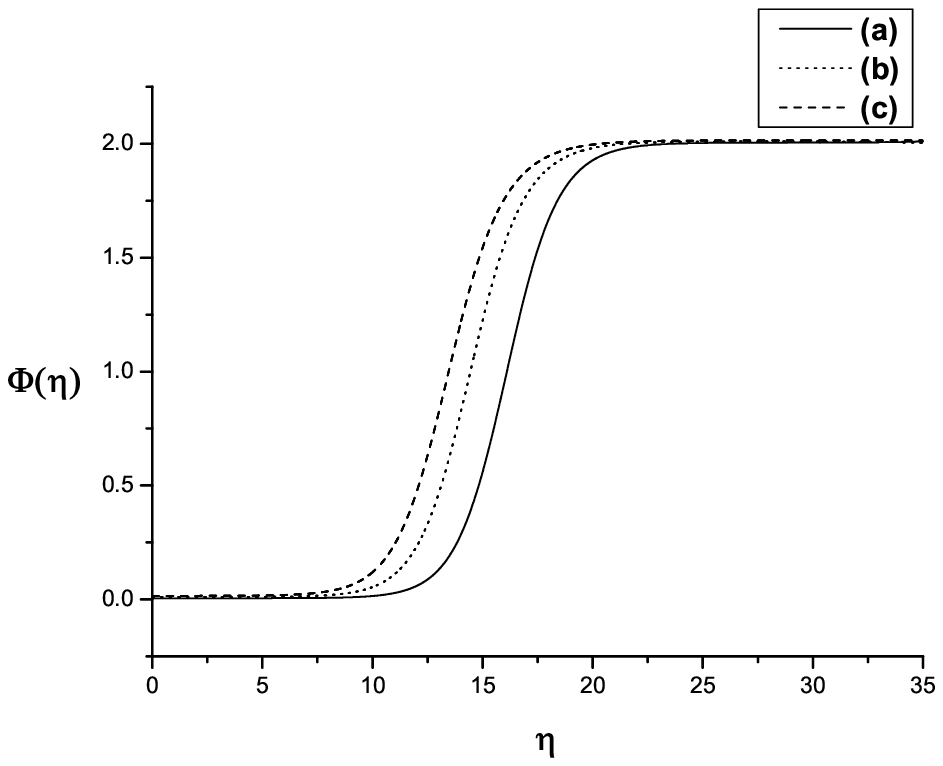}
\end{center}
\caption{\footnotesize{The false vacuum bubble profiles for
several values of $\tilde{\epsilon}$ and $\xi$ in case 2. Here
$\xi$ is taken to be positive. The three curves are (a) solid
curve: $\tilde{\epsilon}=0.01$ and $\xi \simeq 0.311$, (b) dotted
curve: $\tilde{\epsilon}=0.02$ and $\xi \simeq 0.360$, (c) dashed
curve: $\tilde{\epsilon}=0.03$ and $\xi \simeq 0.396$.}}
\label{fig:fig5}
\end{figure}

\begin{figure}[t]
\begin{center}
\includegraphics[width=4.5in]{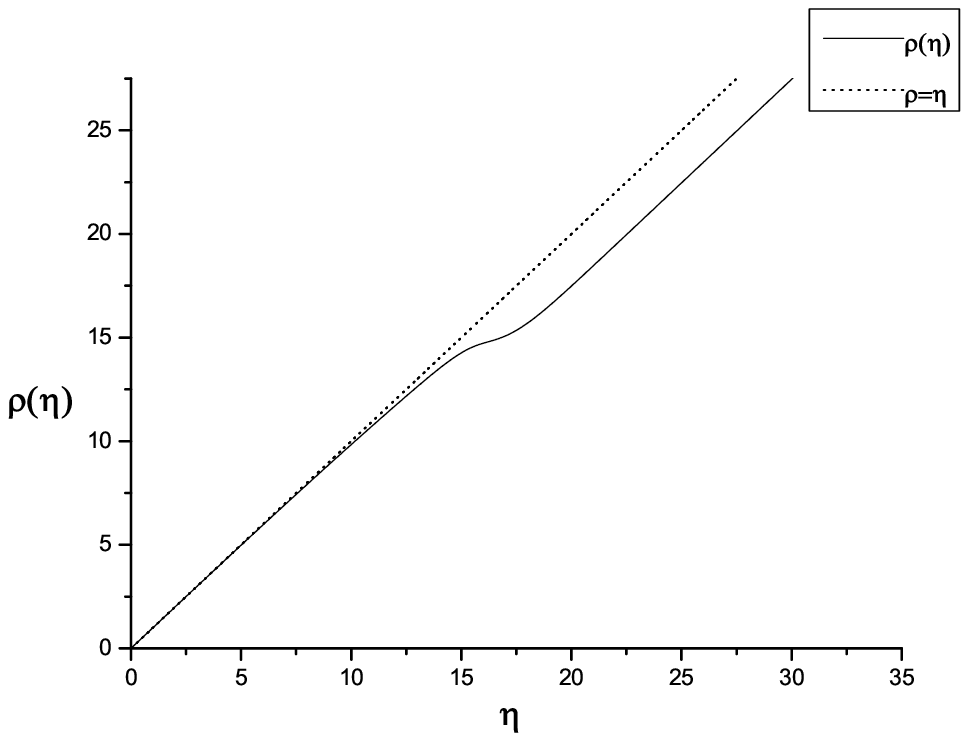}
\end{center}
\caption{\footnotesize{The evolution of
$\tilde{\rho}(\tilde{\eta})$ in case 2. The solid curve is the
solution of $\tilde{\rho}$ with $\tilde{\epsilon} =0.01$. In the
region inside the bubble, $\rho=\Lambda \sin
\frac{\eta}{\Lambda}$, and outside the bubble,
$\rho=\eta+\delta$.}} \label{fig:fig6}
\end{figure}

In Case 2, the true vacuum state with the zero energy density
decays into the false vacuum state with the positive energy
density. We have computed three cases in the similar manner to
Case 1. The corresponding results are presented by (a) solid
curve: $\tilde{\epsilon}=0.01$ and $\xi \simeq 0.311$, (b) dotted
curve: $\tilde{\epsilon}=0.02$ and $\xi \simeq 0.360$, (c) dashed
curve: $\tilde{\epsilon}=0.03$ and $\xi \simeq 0.396$ and we also
take $\tilde{\kappa}=0.3, \tilde{U_T}=0$ (Fig.\ \ref{fig:fig5}).
The solid curve is the solution of $\tilde{\rho}$ with
$\tilde{\epsilon} =0.01$ (Fig.\ \ref{fig:fig6}). In the region
inside the bubble, $\rho=\Lambda \sin \frac{\eta}{\Lambda}$, and
outside the bubble, $\rho=\eta+\delta$.

\begin{figure}[t]
\begin{center}
\includegraphics[width=4.5in]{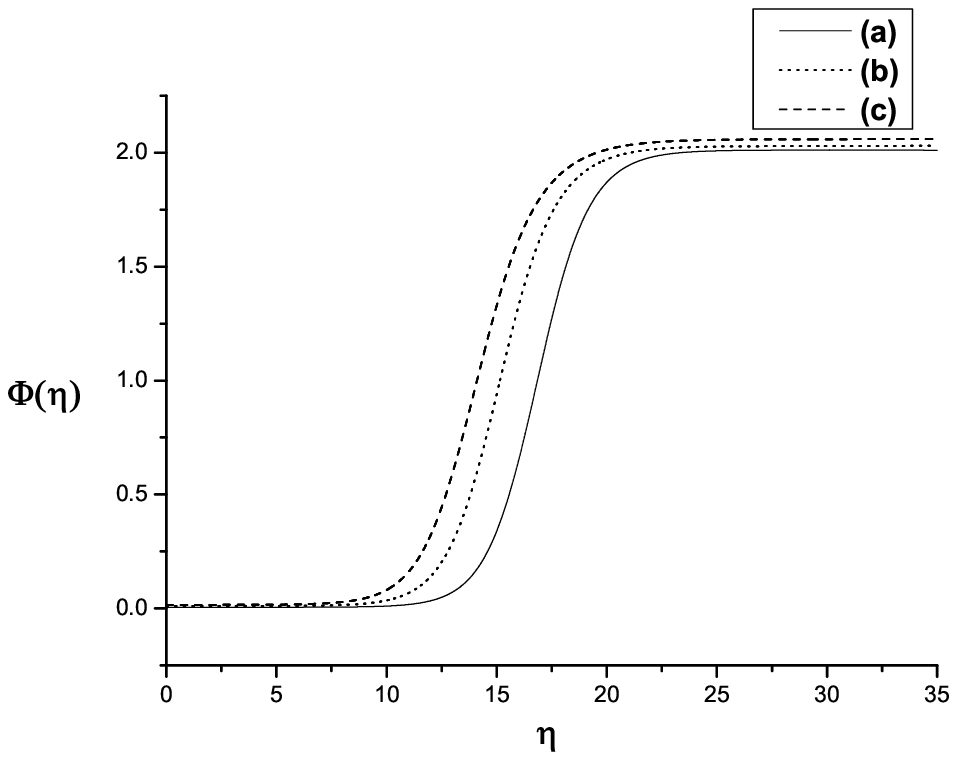}
\end{center}
\caption{\footnotesize{The false vacuum bubble profiles for
several values of $\tilde{\epsilon}$ and $\xi$ in case 3. Here
$\xi$ is taken to be positive. The three curves are (a) solid
curve: $\tilde{\epsilon}=0.01$ and $\xi \simeq 0.328$, (b) dotted
curve: $\tilde{\epsilon}=0.02$ and $\xi \simeq 0.414$, (c) dashed
curve: $\tilde{\epsilon}=0.03$ and $\xi \simeq 0.508$.}}
\label{fig:fig7}
\end{figure}

\begin{figure}[t]
\begin{center}
\includegraphics[width=4.5in]{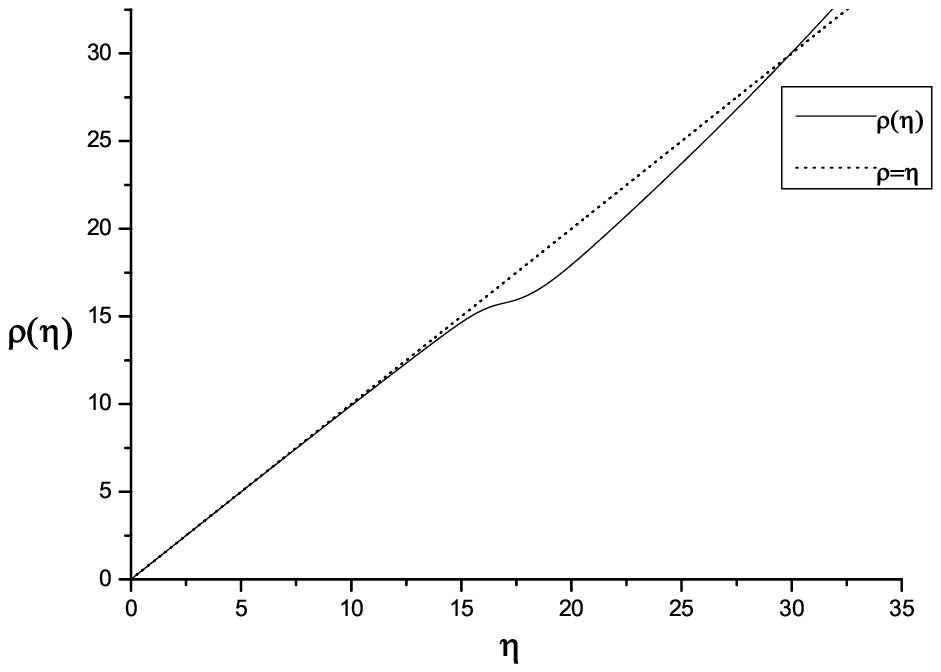}
\end{center}
\caption{\footnotesize{The evolution of
$\tilde{\rho}(\tilde{\eta})$ in case 3. The solid curve is the
solution of $\tilde{\rho}$ with $\tilde{\epsilon} =0.01$. In the
region inside the bubble, $\rho=\Lambda \sin
\frac{\eta}{\Lambda}$, and outside the bubble, $\rho=\Lambda_2
\sinh\frac{\eta+\delta}{\Lambda_2}$.}} \label{fig:fig8}
\end{figure}

In Case 3, the true vacuum state with the negative energy density
decays into the false vacuum state with the positive energy
density. We have computed three cases in the similar manner to
Case 1. The corresponding results are (a) solid curve:
$\tilde{\epsilon}=0.01$ and $\xi \simeq 0.328$, (b) dotted curve:
$\tilde{\epsilon}=0.02$ and $\xi \simeq 0.414$, (c) dashed curve:
$\tilde{\epsilon}=0.03$ and $\xi \simeq 0.508$ and we also take
$\tilde{\kappa}=0.3$ (Fig.\ \ref{fig:fig7}). The solid curve is
the solution of $\tilde{\rho}$ with $\tilde{\epsilon} =0.01$
(Fig.\ \ref{fig:fig8}). In the region inside the bubble,
$\rho=\Lambda \sin \frac{\eta}{\Lambda}$, and outside the bubble,
$\rho=\Lambda_2 \sinh\frac{\eta+\delta}{\Lambda_2}$.

\begin{figure}[t]
\begin{center}
\includegraphics[width=4.5in]{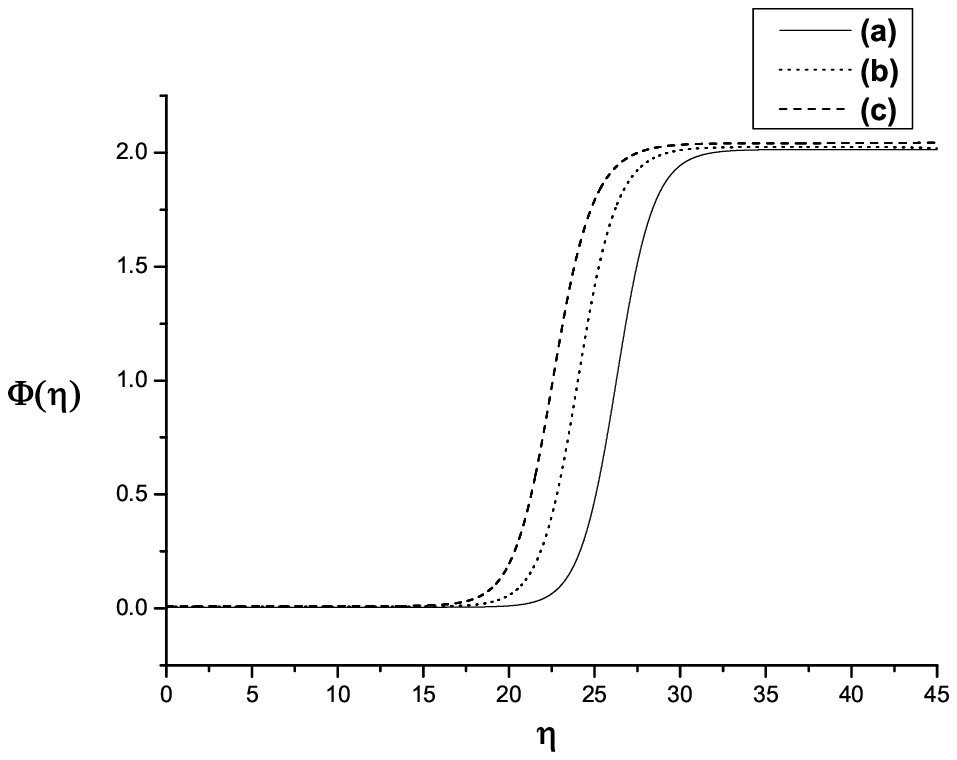}
\end{center}
\caption{\footnotesize{The false vacuum bubble profiles for
several values of $\tilde{\epsilon}$ and $\xi$ in case 4. Here
$\xi$ is taken to be positive. The three curves are (a) solid
curve: $\tilde{\epsilon}=0.01$ and $\xi \simeq 0.258$, (b) dotted
curve: $\tilde{\epsilon}=0.015$ and $\xi \simeq 0.316$, (c) dashed
curve: $\tilde{\epsilon}=0.02$ and $\xi \simeq 0.377$.}}
\label{fig:fig9}
\end{figure}

\begin{figure}[t]
\begin{center}
\includegraphics[width=4.5in]{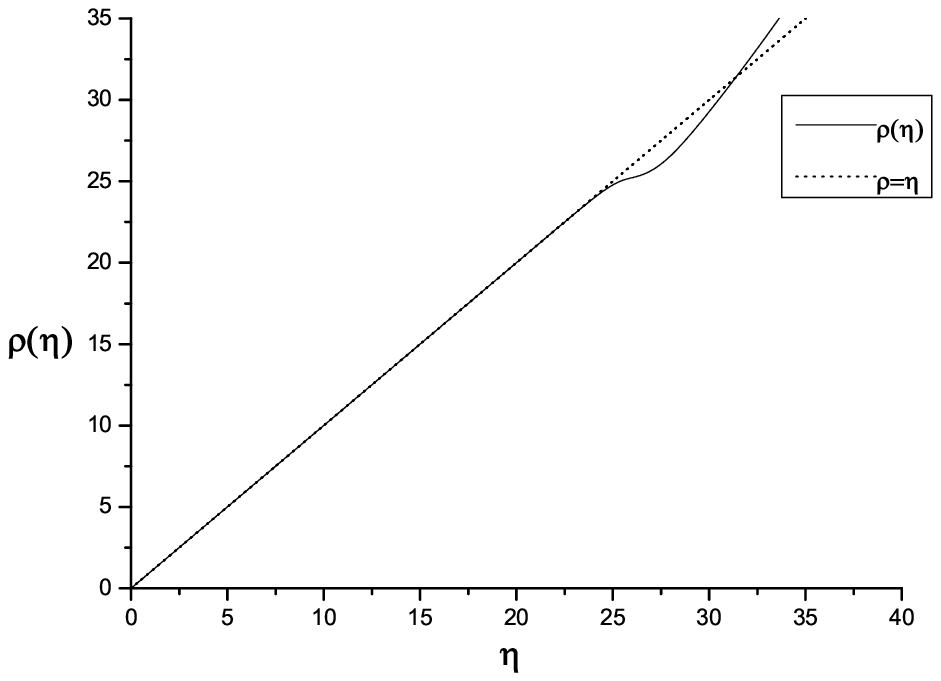}
\end{center}
\caption{\footnotesize{The evolution of
$\tilde{\rho}(\tilde{\eta})$ in case 4. The solid curve is the
solution of $\tilde{\rho}$ with $\tilde{\epsilon} =0.01$. In the
region inside the bubble, $\rho=\eta$, and outside the bubble,
$\rho=\Lambda_2 \sinh\frac{\eta+\delta}{\Lambda_2}$.}}
\label{fig:fig10}
\end{figure}

In Case 4, the true vacuum state with the negative energy density
decays into the false vacuum state with the zero energy density.
The results are (a) solid curve: $\tilde{\epsilon}=0.01$ and $\xi
\simeq 0.258$, (b) dotted curve: $\tilde{\epsilon}=0.015$ and $\xi
\simeq 0.316$, (c) dashed curve: $\tilde{\epsilon}=0.02$ and $\xi
\simeq 0.377$ and we also take $\tilde{\kappa}=0.3, \tilde{U_F}=0$
(Fig.\ \ref{fig:fig9}). The solid curve is the solution of
$\tilde{\rho}$ with $\tilde{\epsilon} =0.01$ (Fig.\
\ref{fig:fig10}). In the region inside the bubble, $\rho=\eta$,
and outside the bubble, $\rho=\Lambda_2
\sinh\frac{\eta+\delta}{\Lambda_2}$.

\begin{figure}[t]
\begin{center}
\includegraphics[width=4.5in]{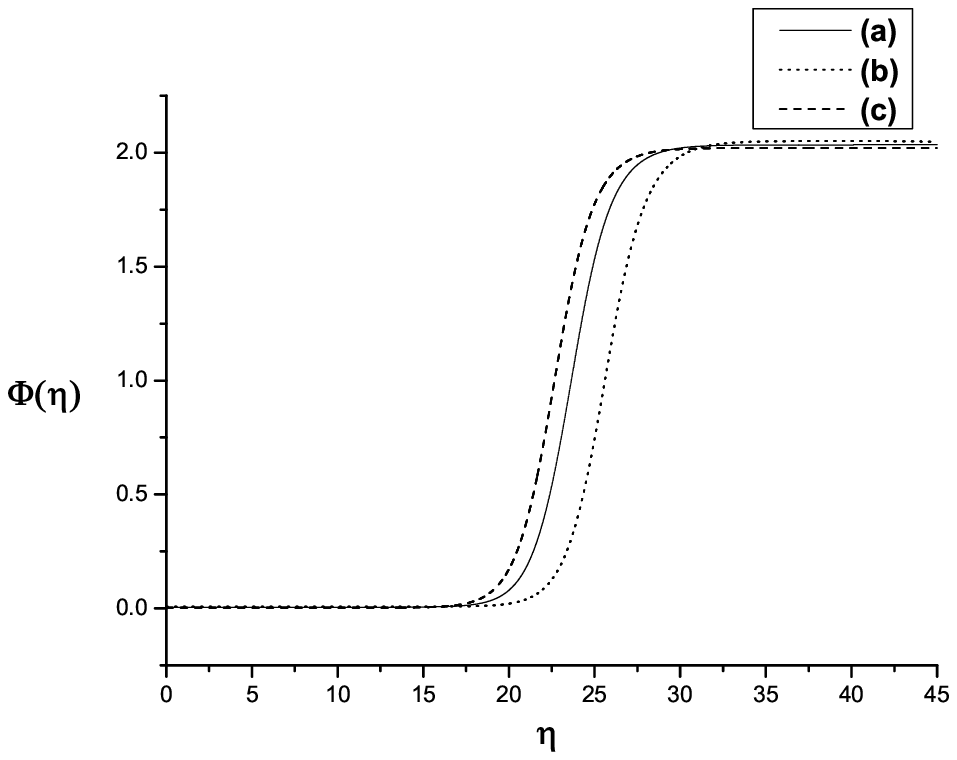}
\end{center}
\caption{\footnotesize{The false vacuum bubble profiles for
several values of $\tilde{\epsilon}$ and $\xi$ in case 5. Here
$\xi$ is taken to be positive. The three curves are (a) solid
curve: $\tilde{\epsilon}=0.01$ and $\xi \simeq 0.359$, (b) dotted
curve: $\tilde{\epsilon}=0.015$ and $\xi \simeq 0.400$, (c) dashed
curve: $\tilde{\epsilon}=0.005$ and $\xi \simeq 0.319$.}}
\label{fig:fig11}
\end{figure}

\begin{figure}[t]
\begin{center}
\includegraphics[width=4.5in]{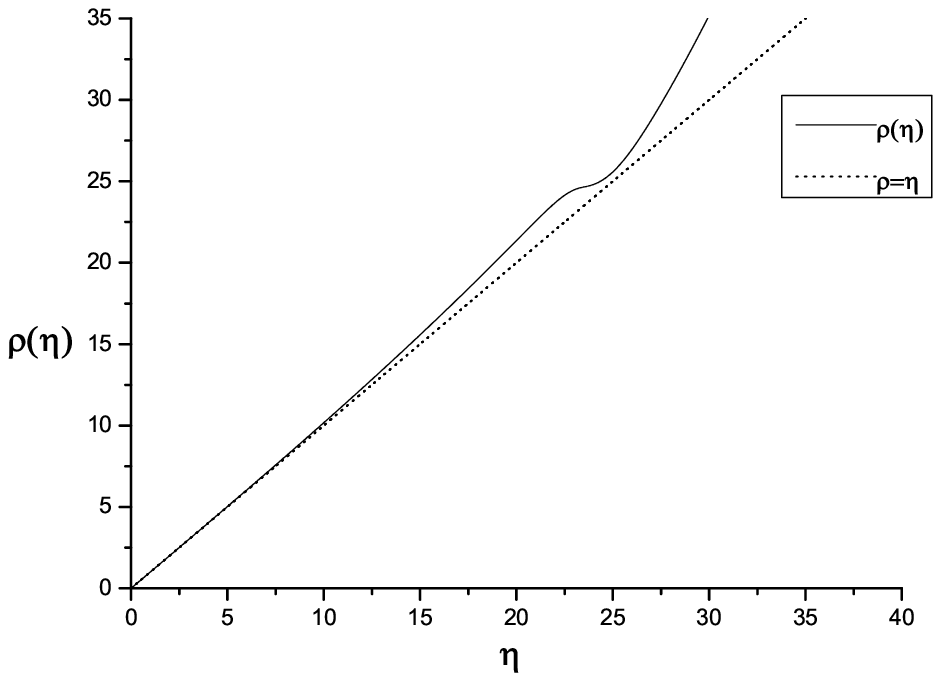}
\end{center}
\caption{\footnotesize{The evolution of
$\tilde{\rho}(\tilde{\eta})$ in case 5. The solid curve is the
solution of $\tilde{\rho}$ with $\tilde{\epsilon} =0.01$. In the
region inside the bubble, $\rho=\Lambda_3 \sinh
\frac{\eta}{\Lambda_3}$, and outside the bubble, $\rho=\Lambda_2
\sinh\frac{\eta+\delta}{\Lambda_2}$.}} \label{fig:fig12}
\end{figure}

In Case 5, the true vacuum state with the higher negative energy
density decays into the false vacuum state with the lower negative
energy density. We have computed three cases in the similar manner
to Case 1. But here we obtain the different results from other
cases. The radius of bubble is diminished as $\tilde{\epsilon}$ is
diminished. The corresponding results are (a) solid curve:
$\tilde{\epsilon}=0.01$ and $\xi \simeq 0.359$, (b) dotted curve:
$\tilde{\epsilon}=0.005$ and $\xi \simeq 0.319$, (c) dashed curve:
$\tilde{\epsilon}=0.015$ and $\xi \simeq 0.400$ and we also take
$\tilde{\kappa}=0.3, \tilde{U_F}=-0.01$ (Fig.\ \ref{fig:fig11}).
The solid curve is the solution of $\tilde{\rho}$ with
$\tilde{\epsilon} =0.01$ (Fig.\ \ref{fig:fig12}). In the region
inside the bubble, $\rho=\Lambda_3 \sinh \frac{\eta}{\Lambda_3}$,
and outside the bubble, $\rho=\Lambda_2
\sinh\frac{\eta+\delta}{\Lambda_2}$.

In summary, we have shown that there exist parameter regions where
the false vacuum bubble can be formed in all cases.

\section{The thin-wall approximation}

In the limit of small $\epsilon$ the field stays near to the top
of the hill of the inverted potential, $-U(\Phi_F)$, for quite a
long time so that $\rho$ grows large with $\Phi$ staying near
$\Phi_F$. As $\rho$ becomes large, the second term becomes
negligible and $\Phi$ quickly goes to $\Phi_T$ and stays at that
point from thereafter. From the Euclidean field equations for
$\Phi$ and $\rho$, the second term on the left hand side of Eq.\
(\ref{ephi}) is given by
\begin{equation}
3\frac{\rho'}{\rho}\Phi'=3 \left[\frac{1}{\rho^2}+
\frac{\kappa(\frac{1}{2}\Phi'^2-U)}{3(1-\xi\Phi^2\kappa)}
\right]\Phi'.
\end{equation}

In this section we assume the thin-wall approximation. The
validity of the thin-wall approximation has been described in
detail by Samuel and Hiscock \cite{samuel}. However, in order to
get the effect of nonminimal coupling we will keep the term in the
Euclidean field equations.

In this approximation, it is justified to neglect the second term,
$\frac{3\rho'}{\rho} \Phi'$, and the bubble nucleation rate is
calculated by $\Gamma /V = Ae^{-B/\hbar}$ where $B$ is the
difference between the Euclidean action of the bounce and that of
the true vacuum state,
\begin{equation}
 B = S^{b}_E - S^{T.V.}_E.
\end{equation}

Thus, the Euclidean action is given by
\begin{eqnarray}
S_E&=&2\pi^2\int_0^{\infty}
d\eta\left[\rho^3\left(\frac{1}{2}(\Phi')^2+U(\Phi)\right)+\frac{3(1-\xi\Phi^2\kappa)}{\kappa}(\rho\rho'^2+\rho^2\rho''-\rho)\right]
\nonumber \\
&=&4\pi^2\int_0^{\infty} d\eta\left[\rho^3U(\Phi) -
\frac{3\rho(1-\xi\Phi^2\kappa)}{\kappa} + 3\xi\rho^2\rho'\Phi\Phi'
\right]. \label{action}
\end{eqnarray}
Here we eliminate the second-derivative term by integration by
parts and use the Euclidean field equation to eliminate $\rho'$.
The third term of the second line in Eq.\ (\ref{action}) is
vanished because of $\rho' \to 0$ in the wall and $\Phi' \to 0$
both inside and outside of the wall, respectively.

Now, we shall use the thin-wall approximation scheme to evaluate
$B$. In this approximation the exponent $B$ can be divided into
three parts
\begin{equation}
 B= B_{in} + B_{wall} + B_{out}.
\end{equation}
Outside the wall,
\begin{equation}
 B_{out}=S_{E}(\Phi_T)-S_{E}(\Phi_T)=0.
\end{equation}
In the wall, we can replace $\rho$ by $\bar{\rho}$ and Eq.\
(\ref{ephi}) can be modified
\begin{equation}
 \frac{d^2\Phi}{d\eta^2} = \frac{dU}{d\Phi}+\xi R_E \Phi.
\end{equation}
Multiplying Eq.\ (\ref{ephi}) by $\frac{d\Phi}{d\eta}$ and then
integrating over $\eta$, one obtains
\begin{equation}
\frac{d\Phi}{d\eta}=-\sqrt{2[U(\Phi)-U(\Phi_T)]+\xi R_E (\Phi^2 -
\Phi_T^2 )},
\end{equation}
where Ricci scalar $R_E$ is a function of $\rho$ only in the wall
and the minus sign is chosen because we are interested in the
region $\Phi_F < \Phi < \Phi_T$. This minus sign is important for
the positive contribution from the wall. We neglect the term
$\int\xi\Phi^2 R_E' d\eta$ in the above equation because we use
$R_E \simeq \frac{6}{\bar{\rho^2}}$ only in the wall in this
thin-wall approximation. In this work we use the condition
$d\Phi/d\eta|_{\Phi_T} =0$, and consider the case where
$\frac{\lambda \epsilon}{\mu^4}$ is small and we approximate the
quantity to the first order of this parameter.

Then, the contribution from the wall, $B_{wall}$, is given by
\begin{eqnarray}
B_{wall}&=&4\pi^2 \bar{\rho^3} \int_{\Phi_F}^{\Phi_T} \left[ (U(\Phi)-U(\Phi_T)) + \frac{3}{\bar{\rho^2}}\xi(\Phi^2-\Phi_T^2)\right]d\eta \nonumber \\
&=& 2\pi^2 \bar{\rho^3}\int_{\Phi_T}^{\Phi_F}
\sqrt{2[U(\Phi)-U(\Phi_T)]+
\frac{6\xi}{\bar{\rho^2}}(\Phi^2 - \Phi_T^2 )}d\Phi \nonumber \\
&=& 2\pi^2 \bar{\rho^3} \left[ \int_{\Phi_T}^{\Phi_F}
\sqrt{2[U(\Phi)-U(\Phi_T)]}d\Phi - \frac{C\xi}{\bar{\rho^2}} \right] \nonumber \\
&=&2\pi^2 \bar{\rho^3}(S_o - \frac{C\xi}{\bar{\rho^2}}),
\label{wall}
\end{eqnarray}
where $S_o = \int_{\Phi_T}^{\Phi_F} \sqrt{2[U(\Phi)-U(\Phi_T)]}
d\Phi$ and $C=\frac{12b}{\sqrt{\lambda}}(1+2
\ln{\frac{4b^4\lambda+\epsilon}{\epsilon}})$. In the third line,
we approximate the quantity in the square root to the first order.
The second term of the fourth line reflects the correction of
Euclidean surface density.

To evaluate the Euclidean action inside the wall, we will use
$d\rho = d\eta [1-\frac{\kappa\rho^2U}{3(1-\xi\Phi^2\kappa)}]$
\begin{eqnarray}
 B_{in} = \frac{12\pi^2}{\kappa^2}\left[
 \frac{(1-\xi\Phi_F^2\kappa)^2\{[1-\frac{\kappa\bar{\rho^2}U(\Phi_F)}{3(1-\xi\Phi_F^2\kappa)}]^{3/2}-1\}}{U(\Phi_F)}\right.&& \nonumber \\
        -\left. \frac{(1-\xi\Phi_T^2\kappa)^2\{[1-\frac{\kappa\bar{\rho^2}U(\Phi_T)}{3(1-\xi\Phi_T^2\kappa)}]^{3/2}-1\}}{U(\Phi_T)}\right]&&.
\end{eqnarray}
This $B_{in}$ is the contribution from inside the bubble.

To find the critical bubble size, $B$ has to be extremized with
respect to $\bar\rho$,
\begin{eqnarray}
\frac{dB}{d\bar\rho} = 12\pi^2 \bar\rho \left[
\left(\frac{1-\xi\Phi_T^2\kappa}{\kappa}\right)\left(1-
\frac{\kappa\bar\rho^2
U(\Phi_T)}{3(1-\xi\Phi_T^2\kappa)}\right)^{1/2} \right.&& \nonumber \\
-\left. \left(\frac{1-\xi\Phi_F^2\kappa}{\kappa}\right)\left(1-
\frac{\kappa\bar\rho^2
U(\Phi_F)}{3(1-\xi\Phi_F^2\kappa)}\right)^{1/2} \right]+6\pi^2
\bar\rho^2 S_o - 2\pi^2 C\xi =0 . &&
\end{eqnarray}
Here we take $\Phi_F =0$, $\Phi_T =2b$,
$\lambda_1^2=[3/\kappa(U_F+U_T)]$ and
$\lambda_2^2=[3/\kappa(U_F-U_T)]$, then the radius of the false
vacuum bubble is given by
\begin{equation}
\bar\rho^2 = \frac{H \pm \sqrt{H^2-ED}}{E}, \label{derho}
\end{equation}
where
$E=[1+2(\frac{\bar\rho_o}{2\lambda_1})^2+(\frac{\bar\rho_o}{2\lambda_2})^4]
+ 8b^2\xi\kappa^3\lambda_{2}^{2}U_T\left( \frac{8b^2\xi
U_T}{3}-\frac{S_o^2}{2}\right)$,
$H=\frac{\bar{\rho_o^2}}{S_o^2}[(2-8b^2\xi\kappa)
\left(\frac{S_o^2}{4}-4b^2\xi U_T
\right)+\frac{\xi(U_F+U_T)}{3}\left(16b^4\xi\kappa-8b^2-\frac{S_o
C\kappa}{6} \right)]$, $D=\frac{\bar{\rho_o^2}}{S_o^2}[\xi
\left(64b^4\xi+\frac{2S_o C}{3}\right)-256b^6\xi^3\kappa -
\frac{8b^2\xi^2S_o C\kappa}{3}]$,
 and $\bar\rho_o=3S_o/[U_F-U_T]$ is the bubble size without
gravity. If $\xi=0$ is substituted into Eq.\ (\ref{derho}), then
the $\bar\rho^2$ is given by
\begin{equation}
\bar\rho_p^2 =
\frac{\bar\rho_o^2}{\left[1+2(\frac{\bar\rho_o}{2\lambda_1})^2+(\frac{\bar\rho_o}{2\lambda_2})^4
\right]}, \label{prho}
\end{equation}
which is consistent with Parke's results \cite{parke}.

In this case the coefficient $B$ is given by
\begin{eqnarray}
B &=& \frac{12\pi^2}{\kappa^2}\left[ \frac{1}{U_F}
\left\{\left(1-\frac{\kappa\bar{\rho^2}U_F}{3}\right)^{3/2}-1
\right\} -\frac{(1-4b^2\xi\kappa)^2}{U_T}
\left\{\left(1-\frac{\kappa\bar{\rho^2}U_T}{3(1-4b^2\xi\kappa)}\right)^{3/2}-1
\right\} \right]
\nonumber \\
&+& 2\pi^2 \bar{\rho^3}\left(S_o-\frac{C\xi}{\bar{\rho^2}}\right),
\end{eqnarray}
here we take the plus sign in Eq.\ (\ref{derho}) and Eq.\
(\ref{antiderho}).

On the other hand, if we consider the true vacuum bubble
nucleation within the false vacuum background in this model, in
the wall
\begin{equation}
B_{1 wall}=2\pi^2 \bar{\rho_1^3}(S_o -
\frac{C_1\xi}{\bar{\rho^2}}),
\end{equation}
where $C_1=\frac{12b}{\sqrt{\lambda}}(1+2
\ln{\frac{\epsilon}{\epsilon-4b^4\lambda}})$. The radius of true
vacuum bubble is given by
\begin{equation}
\bar\rho_1^2 = \frac{H_1 \pm \sqrt{H_1^2-ED_1}}{E},
\label{antiderho}
\end{equation}
where
$E=[1+2(\frac{\bar\rho_o}{2\lambda_1})^2+(\frac{\bar\rho_o}{2\lambda_2})^4]
+ 8b^2\xi\kappa^3\lambda_{2}^{2}U_T\left( \frac{8b^2\xi
U_T}{3}-\frac{S_o^2}{2}\right)$,
$H_1=\frac{\bar{\rho_o^2}}{S_o^2}[(2-8b^2\xi\kappa)
\left(\frac{S_o^2}{4}-4b^2\xi U_T
\right)+\frac{\xi(U_F+U_T)}{3}\left(16b^4\xi\kappa-8b^2-\frac{S_o
C_1\kappa}{6} \right)]$, and $D_1=\frac{\bar{\rho_o^2}}{S_o^2}[\xi
\left(64b^4\xi+\frac{2S_o C_1}{3}\right)-256b^6\xi^3\kappa -
\frac{8b^2\xi^2S_o C_1\kappa}{3}]$ and the coefficient $B_1$ is
given by
\begin{eqnarray}
B_1&=& \frac{12\pi^2}{\kappa^2}\left[
\frac{(1-4b^2\xi\kappa)^2}{U_T}
\left\{\left(1-\frac{\kappa\bar{\rho_1^2}U_T}{3(1-4b^2\xi\kappa)}\right)^{3/2}-1
\right\} -\frac{1}{U_F}
\left\{\left(1-\frac{\kappa\bar{\rho_1^2}U_F}{3}\right)^{3/2}-1
\right\} \right]
\nonumber \\
&+& 2\pi^2
\bar{\rho^3}\left(S_o-\frac{C_1\xi}{\bar{\rho^2}}\right).
\end{eqnarray}
If $\xi=0$ is substituted into the above equation, then the
coefficient $B_1$ is given by
\begin{equation}
B_p=
\frac{2B_o[\{1+(\frac{\bar{\rho_o}}{2\lambda_1})^2\}-\{1+2(\frac{\bar\rho_o}{2\lambda_1})^2+(\frac{\bar\rho_o}{2\lambda_2})^4\}^{1/2}
]}{[(\frac{\bar{\rho_o}}{2\lambda_2})^4
\{(\frac{\lambda_2}{\lambda_1})^2 -1
\}\{1+2(\frac{\bar\rho_o}{2\lambda_1})^2+(\frac{\bar\rho_o}{2\lambda_2})^4\}^{1/2}
]},
\end{equation}
where $B_o=27\pi^2S_o^4/2\epsilon^3$ is the nucleation rate in the
absence of gravity. $B_p$ is obtained by Parke \cite{parke}.

Now we discuss the evolution of the false vacuum bubble after its
nucleation. The false vacuum bubble after its nucleation will
either expand or shrink. Can the false vacuum bubble expand within
the true vacuum background? It's not a trivial problem in curved
spacetime because the energy cannot be globally defined and the
energy conservation does not work for vacuum energy in general.
The dynamics of the false vacuum bubble or inflating regions is
discussed in Refs.\ \cite{blau} by employing the junction
condition. However, in order to progress our analysis
continuously, the spacetime outside bubble will be kept flat
(Minkowski), de Sitter or anti-de Sitter spacetime.

Here we analyze the growth of the bubble, following Ref.\
\cite{chul}, where we have discussed the dynamics of true vacuum
bubble. De Sitter space which includes some aspects pertaining to
bubble nucleation have been described in Ref.\ \cite{lindley}. As
the first step to analyze the growth of the bubble, the Lorentzian
solution is obtained by applying the analytic continuation
\begin{equation}
\chi \to i\chi + \frac{\pi}{2},  \label{analytic}
\end{equation}
to the Euclidean solution. The only difference is that one has to
continue the metric as well as the scalar field, $O(4)$-invariant
Euclidean space into an $O(3,1)$-invariant Lorentzian spacetime.

In case 1, the spacetime metric inside the bubble which is
obtained by applying Eq.\ (\ref{analytic}) to Eq.\
(\ref{inmetric-1}) is
\begin{equation}
ds^2 = d\eta^2 + \Lambda^2 \sin^2 \frac{\eta}{\Lambda} \{ -d\chi^2
+ \cosh^2\chi (d\theta^2 +\sin^2\theta d\phi^2)\},
\label{linmetric-1}
\end{equation}
and the metric outside the bubble which is given by
\begin{equation}
ds^2 = d\eta^2 + \Lambda_1^2 \sin^2 \frac{\eta+\delta}{\Lambda_1}
\{ -d\chi^2 + \cosh^2\chi (d\theta^2 +\sin^2\theta d\phi^2)\},
\label{loutmetric-1}
\end{equation}
which are the spatially inhomogeneous de Sitter like metrics. For
Minkowski spacetime, it becomes the spherical Rindler type
\cite{gerlach}.

We next look for the static spherically symmetric forms of the
metric given in Eq's.\ (\ref{linmetric-1}) to
(\ref{loutmetric-1}). Both inside and outside the bubble, the
coordinate transformation
\begin{eqnarray}
r&=&\Lambda_1 \sin\frac{\eta+\delta}{\Lambda_1}\cosh\chi, \nonumber \\
t&=&\frac{\Lambda_1}{2}\ln
\frac{\cos\frac{\eta+\delta}{\Lambda_1}+\sin\frac{\eta+\delta}{\Lambda_1}\sinh\chi}{\cos\frac{\eta+\delta}{\Lambda_1}-\sin\frac{\eta+\delta}{\Lambda_1}\sinh\chi},
\label{dcort}
\end{eqnarray}
change the metric to
\begin{eqnarray}
ds^2&=&- \left(1- \frac{r^2}{\Lambda^2} \right)dt^2 +
\frac{dr^2}{1-\frac{r^2}{\Lambda^2}} +r^2(d\theta^2 + \sin^2\theta
d\phi^2), \nonumber \\
ds^2&=&- \left(1- \frac{r^2}{\Lambda_1^2} \right)dt^2 +
\frac{dr^2}{1-\frac{r^2}{\Lambda_1^2}} +r^2(d\theta^2 +
\sin^2\theta d\phi^2),
\end{eqnarray}
respectively, which are the static spherically symmetric forms of
the de Sitter metric. The region of the bubble wall cannot be
represented by a static spherically symmetric metric. However,
when the wall width is very small so that the thin wall
approximation is valid, almost all of the entire space except the
thin bubble wall region can be approximately represented by static
spherically symmetric metrics. The middle of the bubble wall can
be considered to be located at some constant value of $\eta$, say
$\bar{\eta}$, in the coordinate system of Eq.\ (\ref{gemetric}).
Now let us consider a spacetime point corresponding to a constant
value of $\eta$, say $\eta_c$, which is close to $\bar{\eta}$, but
still within the region that can approximately be represented by a
static spherically symmetric metric. Tracing the motion of this
point is then almost the same as tracing the motion of the bubble
wall. We now proceed to trace the motion of such a point just
inside or outside the bubble wall. For a point just inside the
bubble wall in case 1, we differentiate Eq.\ (\ref{dcort}) with
respect to $\chi$ keeping $\eta$ constant as $\eta=\eta_c$ to
obtain
\begin{eqnarray}
dr&=&\Lambda_1 \sin\frac{\eta_c+\delta}{\Lambda_1}\sinh\chi d\chi, \nonumber \\
dt&=&\Lambda_1 \frac{\sin\frac{\eta_c+\delta}{\Lambda_1}\cos
\frac{\eta_c+\delta}{\Lambda_1}\cosh\chi}{1-\sin^2
\frac{\eta_c+\delta}{\Lambda_1}\cosh^2 \chi } d\chi, \nonumber \\
d\tau&=& \pm
\sqrt{(1-\frac{r^2}{\Lambda_1^2})dt^2-\frac{dr^2}{(1-\frac{r^2}{\Lambda_1^2})}}.
\end{eqnarray}
The growth rate of the bubble wall radius per unit proper time
seen from outside the wall is then calculated to be approximately
\begin{equation}
\frac{dr}{d\tau} = \sqrt{\frac{r^2}{r_o^2}-1},
\end{equation}
where $r_o=\Lambda_1\sin\frac{\eta_c+\delta}{\Lambda_1}$ is the
value of $r$ at $t=0 (\chi=0)$. Here we take the positive value
because the quantity must be the positive value in the square
root, that is to say, it represents the expansion of the false
vacuum bubble and also repulsive nature \cite{ipser}. The $d\tau$
goes to zero as the bubble wall becomes large, so the proper
velocity has a large value. We can also obtain the same form
inside the bubble wall. Furthermore, we can obtain the same form
in every other cases (see Ref.\ \cite{chul}). In future work we
will discuss further the dynamics of the false vacuum bubble by
employing the junction condition.

\section{Summary and Discussions}

In this paper we have shown that the false vacuum bubble can be
nucleated within the true vacuum background in the Einstein theory
of gravity with a nonminimally coupled scalar field by solving the
Euclidean field equations of motion in the semiclassical
approximation.

In section 2 we have presented the formalism for the false vacuum
bubble nucleation within the true vacuum background and our main
idea for this work.

In section 3 we have considered the true-to-false vacuum phase
transitions in the five particular cases; (Case 1) from de Sitter
space to de Sitter space, (Case 2) from flat space to de Sitter
space, (Case 3) from anti-de Sitter space to de Sitter space,
(Case 4) from anti-de Sitter space to flat space and (Case 5) from
anti-de Sitter space to anti-de Sitter space. We have obtained
numerical solutions. In case 1, we have obtained the false vacuum
smaller than the true vacuum horizon. The case 3 can be
interesting from the aspect of the so-called string landscape
\cite{susskind}, which has included both anti-de Sitter and de
Sitter minima \cite{kachru}. Our solution represent how the false
vacuum bubble, corresponding to the de Sitter spacetime, can be
nucleated within the true vacuum background, corresponding to the
anti-de Sitter spacetime.

The thick bubble wall exists when the difference in energy density
between the false and true vacuum is large. For the infinitely
thick bubble wall, the Hawking-Moss transition has been known
\cite{hawking} and discussed in Refs.\ \cite{jensen} although the
transition is certainly not evident from the thin-wall study of
Coleman and De Luccia.

In section 4 we obtained the exponent $B$ and the radius of the
false vacuum bubble approximately by employing Coleman's thin-wall
approximation in the cases both the false vacuum bubble nucleation
within the true vacuum background and the true vacuum bubble
nucleation within the false vacuum background.

If the vacuum-to-vacuum phase transitions occur one after another,
false-to-true and true-to-false, the whole spacetime will have the
complicated vacuum or spacetime structure like as an onion with
different number of the coats everywhere. The black hole creation
\cite{bousso} in our model may make the whole spacetime more
chaotic one.

In addition, using the analytic continuation we have discussed the
evolution of the false vacuum bubble in Lorentzian spacetime. Here
we have shown that the proper circumferential radius of the bubble
wall grows according to $\frac{dr}{d\tau} =
\sqrt{\frac{r^2}{r_o^2}-1}$ both inside and outside the wall in
thin-wall approximation. The $d\tau$ goes to zero as the bubble
wall becomes large.

In fact, in order to analyze more precisely the spacetime outside
the bubble, we need to take the Schwarzschild or Schwarzschild-de
Sitter (anti-de Sitter) spacetime \cite{kottler} by Birkhoff's
theorem \cite{birkhoff} although one can not obtain the
Schwarzschild or Schwarzschild-de Sitter (anti-de Sitter)
spacetime from Eq.\ (\ref{outmetric-1}), (\ref{outmetric-2}),
(\ref{outmetric-3}) by coordinate transformations. The dynamics of
the false vacuum bubble or inflating regions is discussed in
Refs.\ \cite{blau} by employing the junction condition.

In summary, we conclude that the false vacuum bubble can be
nucleated within the true vacuum background due to the term, $-\xi
R_E\Phi$, in the Einstein theory of gravity with a nonminimally
coupled scalar field and expect the phenomenon can be possible in
many other theory of gravities with similar terms.

\section*{Acknowledgements}

We would like to thank Kimyeong Lee, Piljin Yi, Ho-Ung Yee at the
Korea Institute for Advanced Study for their hospitality and
valuable discussions and Yoonbai Kim, Won Tae Kim, Mu-In Park, Hee
Il Kim and Seoktae Koh for stimulating discussions and kind
comments. This work was supported by the Science Research Center
Program of the Korea Science and Engineering Foundation through
the Center for Quantum Spacetime (CQUeST) of Sogang University
with grant number R11 - 2005- 021, Korea Research Foundation grant
number C00111, and supported by the Sogang University Research
Grants in 2004.

\end{document}